\documentclass[conference, 12pt]{IEEEtran}
\IEEEoverridecommandlockouts
\usepackage{cite}
\usepackage{amsmath,amssymb,amsfonts}
\usepackage{algorithmic}
\usepackage{graphicx}
\usepackage{tikz}
\usepackage{caption}
\usepackage{subcaption}
\usepackage{textcomp}
\usepackage{lipsum}
\usepackage{float}
\usepackage{xcolor}
\def\BibTeX{{\rm B\kern-.05em{\sc i\kern-.025em b}\kern-.08em
    T\kern-.1667em\lower.7ex\hbox{E}\kern-.125emX}}
\begin{document}

\title{Effects of Non-Cognitive Factors on Post-Secondary Persistence of Deaf Students: An Agent-Based Modeling Approach}

\author{\IEEEauthorblockN{Marie Alaghband$^1$, Ivan Garibay$^2$}
\IEEEauthorblockA{$^1$\texttt{marie.alaghband@knights.ucf.edu}\\
$^2$\texttt{igaribay@ucf.edu}\\
$^{1,2}$\textit{Department of Industrial Engineering and Management Systems} \\
\textit{University of Central Florida}\\
Orlando, Florida, USA}
}

\maketitle

\begin{abstract}
Post-secondary education persistence is the likelihood of a student remaining in post-secondary education. Although statistics show that post-secondary persistence for deaf students has increased recently, there are still many obstacles obstructing students from completing their post-secondary degree goals. Therefore, increasing the persistence rate is crucial to increase education and work goals for deaf students. In this work, we present an agent-based model using NetLogo software for the persistence phenomena of deaf students. We consider four non-cognitive factors: having clear goals, social integration, social skills, and academic experience, which influence the departure decision of deaf students. Progress and results of this work suggest that agent-based modeling approaches promise to give better understanding of what will increase persistence.
\end{abstract}


\section{Introduction}\label{intro}
Post-secondary persistence refers to the likelihood of student retention in post-secondary education (e.g., university, collage), especially after the first year of enrollment. Since retention in post-secondary education affects college students in many different aspects \cite{b1}, it has received considerable attention in the last five decades \cite{b2}. Students' retention, program completion, and graduation, advances the overall quality of life for people with and without disabilities \cite{b3}. Students with disabilities may have sensory, mobility, mental, emotional or cognitive disabilities. Because of these disabilities, disabled students often encounter more barriers than other students and they complete post-secondary education at lower rates \cite{b4}.

Deafness or severe hearing impairment is considered as a kind of sensory impairment and a disability \cite{b5}. Compared to the general student population, deaf students find the transition to post-secondary setting more problematic \cite{b6, b7}. Based on the National Deaf Center's (NDC) most recent report, about $1.3\%$ of all currently enrolled college students are deaf \cite{b8}. Although post-secondary enrollment rates for deaf people have increased since the 1980s, the completion degree college rate is still fewer than their hearing peers \cite{b9}. These statistics show that there are many deaf students who face obstacles preventing them from completing their post-secondary degree goals. Therefore, increasing the persistence rate in these students plays an integral role in increasing education and work goals of deaf students. 

Much research has been conducted to identify and study factors affecting the post-secondary enrollment, persistence, completion, and graduation rates for deaf students \cite{b10, b11, b12, b1, b13, b14, b15, b32}. These studies show that cognitive factors such as academic preparation and English literacy are the important in post-secondary enrollment and success rate predictions of deaf students \cite{b5}, but not for completion or graduation. These findings indicate that deaf students with adequate academic skills are still likely to drop out of college \cite{b16}. Hence, after deaf students enrol; in a post-secondary institution, other factors such as personal and non-cognitive factors are considered as stronger predictors for academic persistence and graduation rates \cite{b17, b18, b19, b20}. Some important non-cognitive factors influencing post-secondary persistent of deaf students are considered as academic experience, social integration, social skills, and clear goals and strategies. 

Positive academic experiences such as having informal mentorship from faculty, participating in college activities outside of class, and collaborating with academic advisor, as well as high levels of social integration, such as being satisfied with social experience and having the ability to adjust socially, have a direct influence on students who persist post-secondary education \cite{b21, b4}. Another important non-cognitive factor is social skills which include a high level of involvement in social activities, and the ability to perceive social situations and respond to the behaviors of others \cite{b22, b11}. Last but not least, having clear goals and strategies helps deaf students to have self-confidence and the desire to overcome post-secondary barriers \cite{b23}.

Agent-based modeling (ABM) is a computational method that allows us to create, analyze, and model a system composed of autonomous decision-making artificial entities called agents \cite{b24}. 
ABMs are usually used in cases of modeling real-world phenomena that need more generalized models which can adapt to our world. ABMs can be coupled with other well developing methods such as machine learning --an area of artificial intelligence that attracted attentions in various fields of research such as cyber security \cite{b33} and computer vision \cite{b34}-- to alter and enhance the way we analyze all different kinds of data.

An agent in ABM is an artificial autonomous individual who has properties, actions, and goals, which enables it to assess situations and make decisions based on defined rules \cite{b25}. In ABMs, agents may have interactions with themselves, other agents, and/or environments that permits us to execute and study how rules of agent behavior give rise to the emergence of macro-phenomena as the simulation output \cite{b26}. This capability of ABMs in capturing emergent macro-phenomena, along with other benefits in providing a natural description of a system and flexibility, has made them a popular modeling approach in various fields of research \cite{b27}.

Deaf student post-secondary persistence is affected by various factors which if modified, can lead to a better life by providing the necessary boost for education and employment goals \cite{b10,b1,b32}. Such important phenomena can be best understood by using a bottom-up approach; ABM. Despite previously conducted research which identified non-cognitive factors influencing the post-secondary persistence of deaf students, interactions between these factors and their influence in predicting the persistence and graduation rate is still a barely explored field of research. In this work, we present an agent-based computational model for post-secondary persistence of deaf students with the goal of studying the effects of non-cognitive factors in post-secondary persistence. Based on the literature, academic experience, social integration, social skill, and clear goals and strategies described before are considered as four non-cognitive factors influencing post-secondary persistence for deaf students. To the best of our knowledge, this is the first agent-based modeling simulation for measuring the influence of non-cognitive factor on post-secondary persistence in deaf students. The contributions of this work include presenting the first agent-based modeling simulation of the effects of non-cognitive factors in deaf student post-secondary persistence rate.

\section{Methodology}\label{method}
\begin{figure*}[t]
    \centering
    \includegraphics[width=\textwidth,height=9cm,width=13cm, scale=0.7]{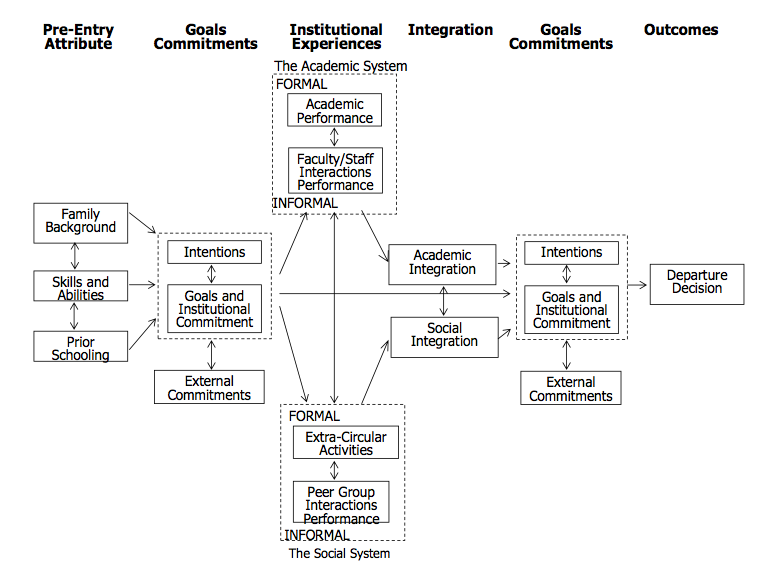}
    \caption{A conceptual diagram for dropout from college presented by \cite{b28}}
    \label{Tinto model}
\end{figure*}

\begin{figure*}
    \centering
    \includegraphics[width=\textwidth,height=4cm,width=12cm, scale=0.7]{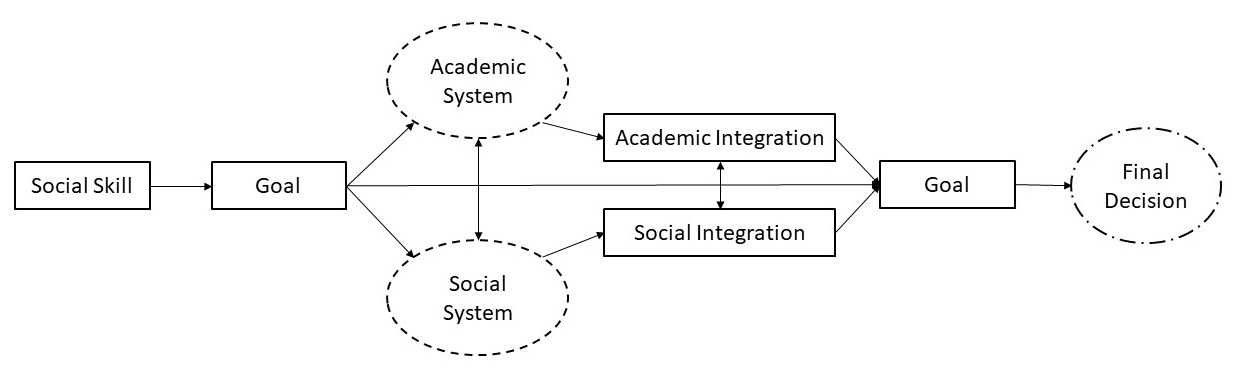}
    \caption{Diagram of four non-cognitive factors effecting student's departure decision based on Tinto's dropout from college model}
    \label{my model}
\end{figure*}

Among several student retention models, the one presented by Tinto \cite{b28, b29} is held in high regard and is the most cited model \cite{b30, b2}. Tinto's model, shown in Figure \ref{Tinto model}, provides a heuristic and theoretical framework for understanding student behaviour while describing the factors influencing the persistence process. With some modifications, this model can be applied to deaf college students as well \cite{b10, b1}. According to this theory, a combination of student characteristics and academical, environmental, and social integration in an institution influence the student's departure decision. To create our model, we use the theoretical model of Tinto \cite{b29}. By only considering the influence of four non-cognitive factors (academic experience, social integration, social skill, and clear goals) on Tinto's model, we created a new simplified framework as shown in Figure\ref{my model}. This figure shows the diagram of four non-cognitive factors influencing student departure decision utilizing the same relations as Tinto's model (Figure\ref{Tinto model}). In this work, this theoretical framework is used to create the ABM and measure the probability of the departure decision of a student.

Our proposed ABM model involves two categories of actors: teachers and deaf agents. Deaf agents are members of the general population who have graduated from secondary school and are capable of attending post-secondary education regardless of their age. While teachers stay at college during college years (runs of the model), deaf agents can decide whether to attend college or not. The specifications, attributes, and behavioral rules of teachers and deaf agents during the SETUP and RUN phases of the ABM are as follows;

\textbf{SETUP Specifications.} In order to setup the model, the user sets exogenous parameters of the total number of agents available in the model (called "\textit{num\_agents}"), fraction of teachers (called "\textit{frac\_teachers}"), and college attendance percentage of deaf students (called "\textit{College\_Attendance}"). The total number of agents will show the sum of teachers and deaf agents. To show the mathematical equations, the total number of teachers and deaf agents is equal to:

\small
$$num\_teachers = num\_agents * frac\_teachers$$
$$num\_deaf\_agents = num\_agents - num\_teachers$$\\

\normalsize

\begin{figure*}[]
\begin{subfigure}{0.5\textwidth}
    \centering
    \includegraphics[width=\linewidth,height=6cm,width=6cm, scale=0.5]{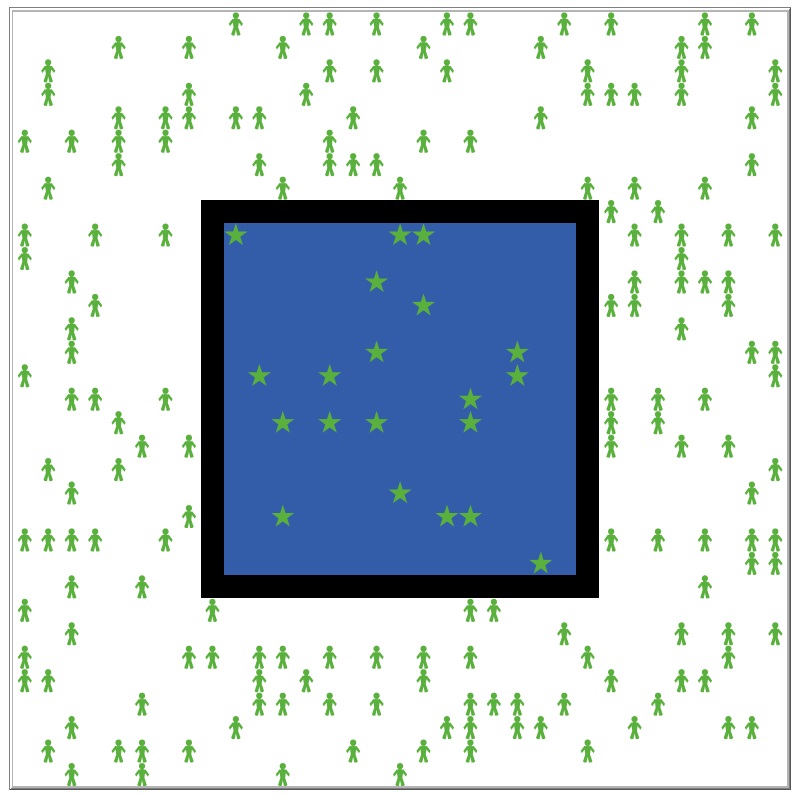}
    \caption{Setup screen of the student retention ABM. The blue square in the middle of the screen with a thick black edge shows the college. White locations surrounding the college are considered residential locations. The green human shapes on the residential locations and the green stars at college represent deaf agents and the teachers, respectively.}
    \label{setup}
\end{subfigure}\hspace*{\fill}
\begin{subfigure}{0.5\textwidth}
    \centering
    \includegraphics[width=\linewidth,height=6cm,width=6cm, scale=0.5]{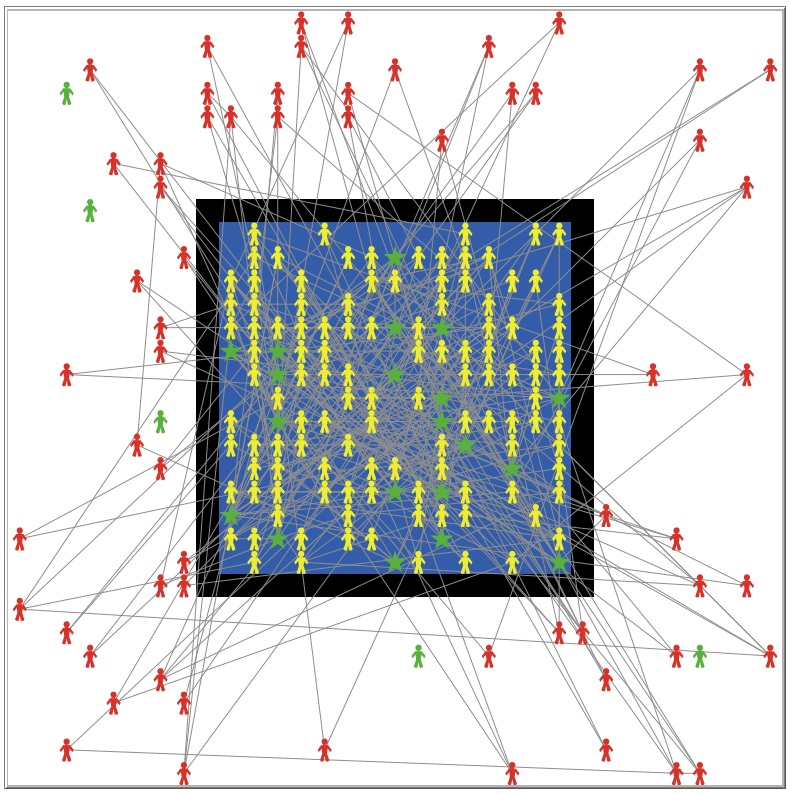}
    \caption{Screen of the student retention ABM while running the model. As soon as deaf agents attend college at the first tick (year), their color changes to yellow and they will be called students. If students quit the college, their color turns to red and they will leave college and move into a random location on the residential locations. The gray lines connecting agents show the created links of each student with teachers and other students.}
    \label{during}
\end{subfigure}\hspace*{\fill}

\medskip
\centering
\begin{subfigure}{0.5\textwidth}
    \centering
    \includegraphics[width=\linewidth,height=6cm,width=6cm, scale=0.5]{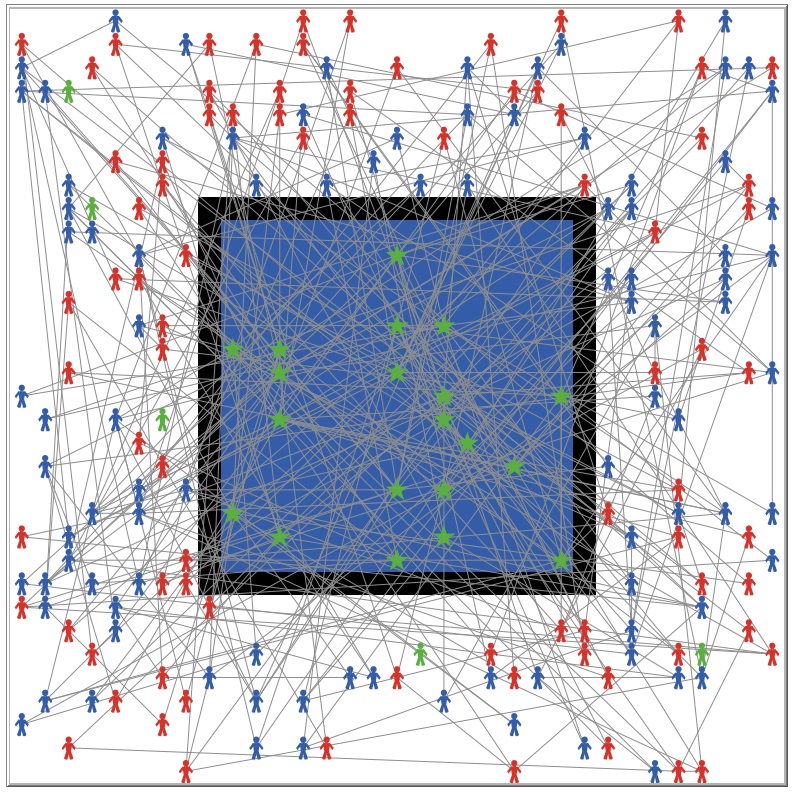}
    \caption{Screen of the student retention ABM after four ticks. At the end of the fourth year, students will graduate. By graduating, their color will turn to blue and they will move into one of the residential locations (leave college).}
    \label{end}
\end{subfigure}
\caption{Exemplary visualizations of the proposed NetLogo model at setup, after two ticks, and after four ticks of running the model.}
\label{screens}
\end{figure*}

In the model settings, academic experience, social integration, social skill, and clear goals are named as \textit{Academic\_Experience}, \textit{Social\_Integration}, \textit{Social\_Skill}, and \textit{Goal}, respectively. Clear goals and social skill factors are considered to be exogenous and heterogeneous across agents. Lacking real-world data, each agent's value for the two exogenous factors are considered to have the uniform distribution on the interval of (0,1); $U(0,1)$. Setting the value of $0$ for each of these factors show the lowest level of the factor. As the value increases, the level of having that positive factor increases, while "$1$" is the highest level of the factor.

On the other hand, the other two factors, academic experience and social integration, (which are usually dependent to connections and interactions with others, such as teachers and other students at the college), 
are considered to be derived from a multiplication of the initial value of the factor that the user sets and the number of the links (connections) with others that a student create at college.


\begin{table*}[t]
\caption{Input assumptions for all runs and experiments. Factors values changes with steps of 0.1 in the interval of (0.1,1).}
\label{runs}
\resizebox{\textwidth}{!}{%
\begin{tabular}{l|cclcccc}
factor influence & num\_agents & college\_attendance & repetitions & goal & academic experience & social skill & social integration \\ \hline
goal & 200 & 87.2 & 10 & (0.1,1) & 0.5 & 0.5 & 0.5 \\
academic experience & 200 & 87.2 & 10 & 0.5 & (0.1,1) & 0.5 & 0.5 \\
social skill & 200 & 87.2 & 10 & 0.5 & 0.5 & (0.1,1) & 0.5 \\
social integration & 200 & 87.2 & 10 & 0.5 & 0.5 & 0.5 & (0.1,1)
\end{tabular}%
}
\end{table*}

\begin{figure*}[t]
\begin{subfigure}{0.5\textwidth}
    \centering
    \includegraphics[width=\linewidth,height=5cm,width=6.5cm, scale=0.5]{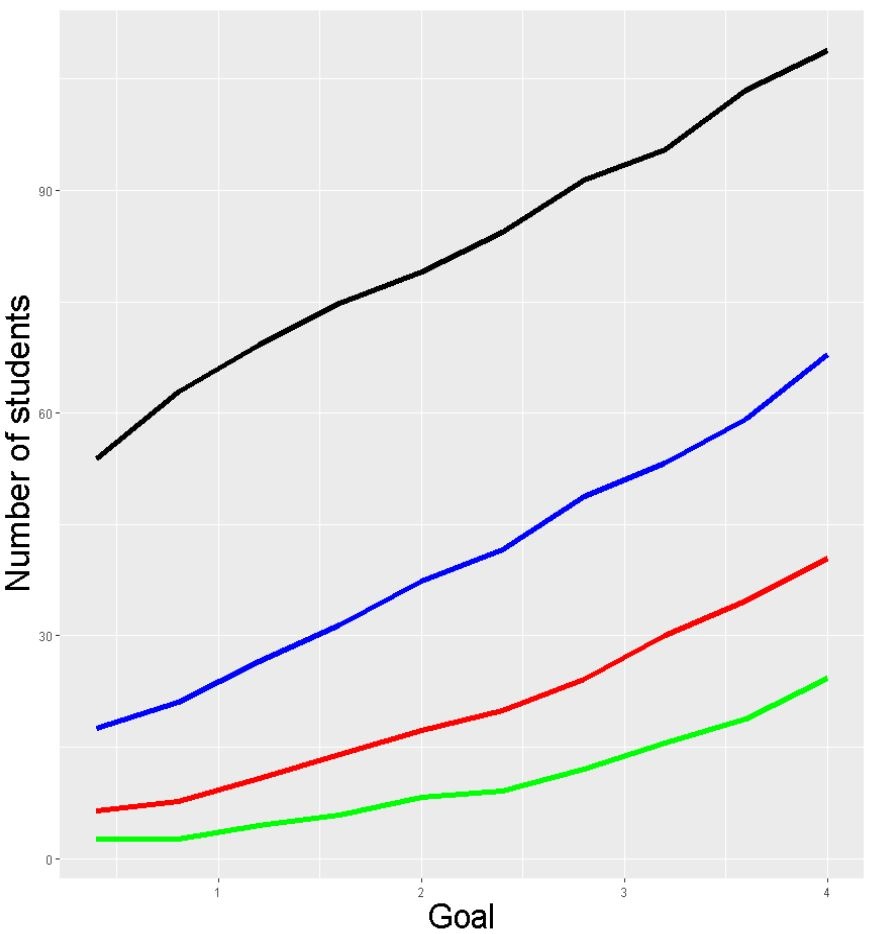}
    \caption{Goal factor}
    \label{goal}
\end{subfigure}\hspace*{\fill}
\begin{subfigure}{0.5\textwidth}
    \centering
    \includegraphics[width=\linewidth,height=5cm,width=9cm, scale=0.5]{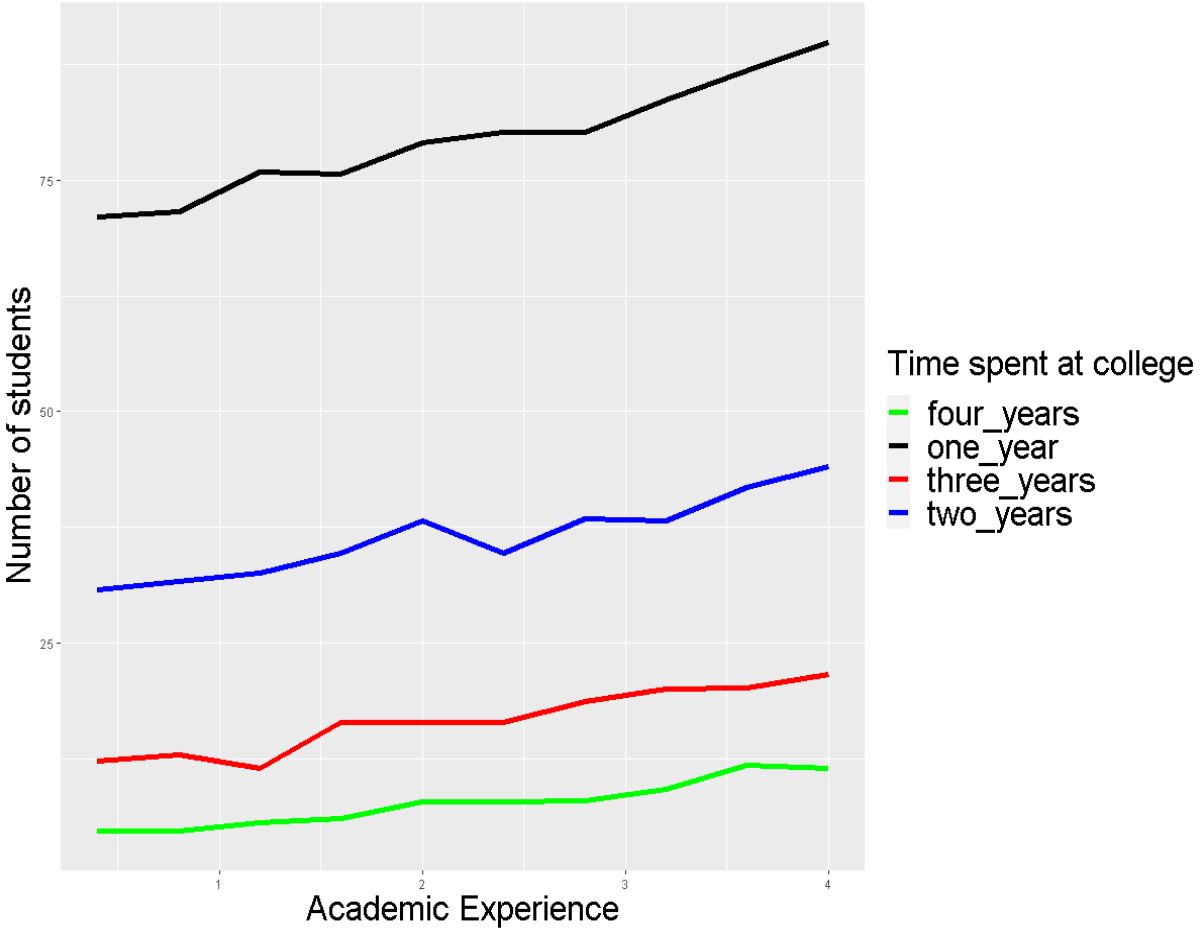}
    \caption{Academic experience factor}
    \label{academic_experience}
\end{subfigure}\hspace*{\fill}

\medskip

\centering
\begin{subfigure}{0.5\textwidth}
    \centering
    \includegraphics[width=\linewidth,height=5cm,width=6.5cm, scale=0.5]{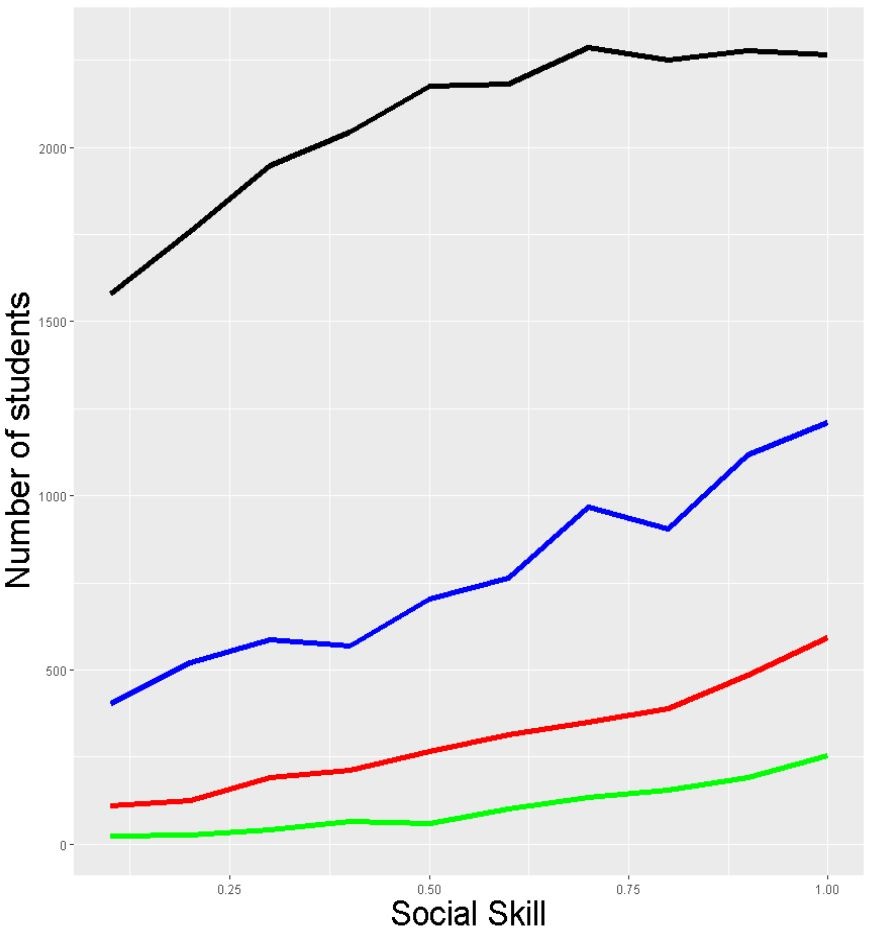}
    \caption{Social skill factor}
    \label{social_skill}
\end{subfigure}\hspace*{\fill}
\begin{subfigure}{0.5\textwidth}
    \centering
    \includegraphics[width=\linewidth,height=5cm,width=9cm, scale=0.5]{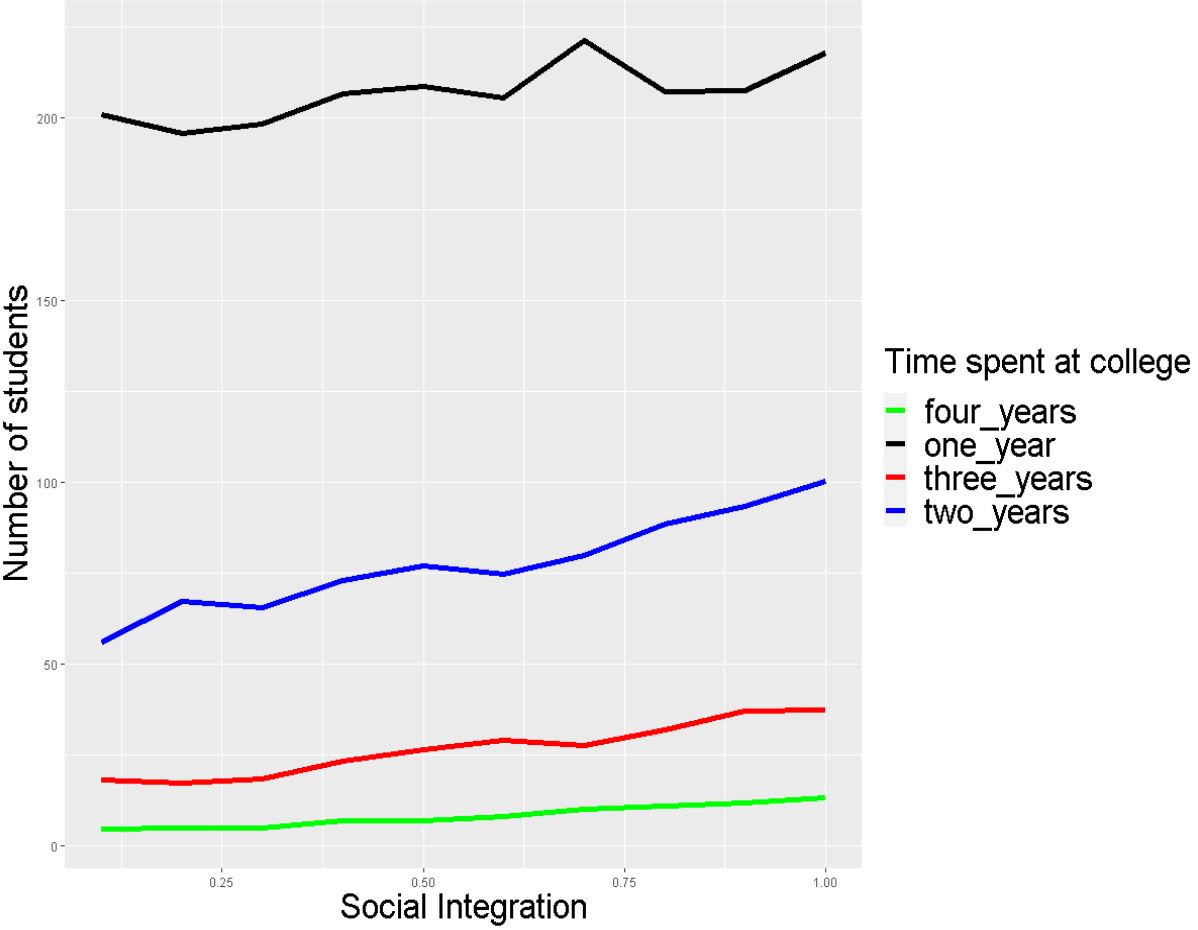}
    \caption{Social integration factor}
    \label{social_integration}
\end{subfigure}\hspace*{\fill}
\caption{Results of behavior space experiment for four non-cognitive factors of goal, academic experience, social skill, and social integration. Each graph shows the result of behavior search for one factor changing in an interval of (0,1) with steps of $0.1$.  To run the experiment for one factor, all other three factors are considered to have a value of $0.5$.}
\label{years}
\end{figure*}

There are two locations in our ABM screen: college, and residential. College is an institution that deaf agents can attend to pursue their post-secondary education. College is located at the middle of the screen and is colored in blue with a thick black line around outside edges. On the other hand, residential locations can be any other kind of location in which deaf people cannot have post-secondary education. Residential locations are represented by the white color surrounding the college. Figure \ref{setup} shows both college and residential locations. When the model starts, only teachers are located at the college; all deaf agents are randomly located in one of the empty spaces of the residential locations. Figure \ref{setup} shows the setup screen of the proposed ABM. In Figure \ref{setup}, deaf agents are shown as green human shapes located at white residential locations while teachers are shown as green stars located randomly in the college.

\begin{figure*}[hbt!]
\begin{subfigure}{0.45\textwidth}
    \centering
    \includegraphics[width=\linewidth,height=5cm,width=6.5cm, scale=0.5]{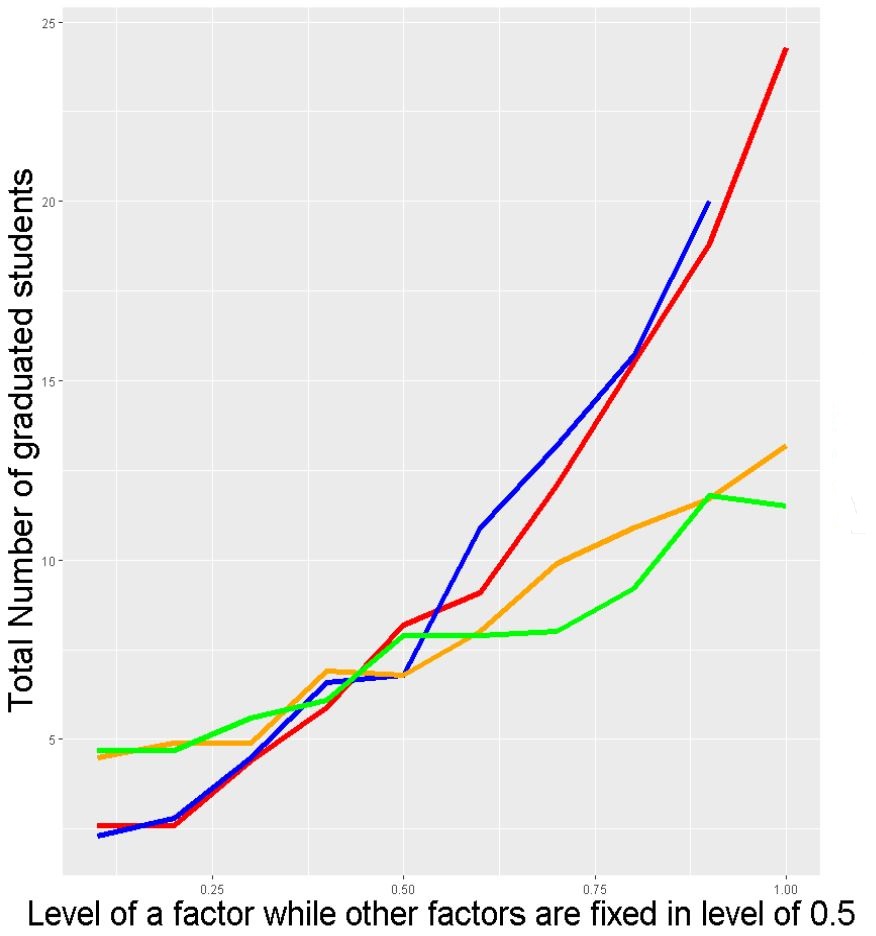}
    \caption{Number of graduated students based on different levels of one factor while fixing the other factors.}
    \label{graduated}
\end{subfigure}\hspace*{\fill}
\begin{subfigure}{0.45\textwidth}
    \centering
    \includegraphics[width=\linewidth,height=5cm,width=9cm, scale=0.5]{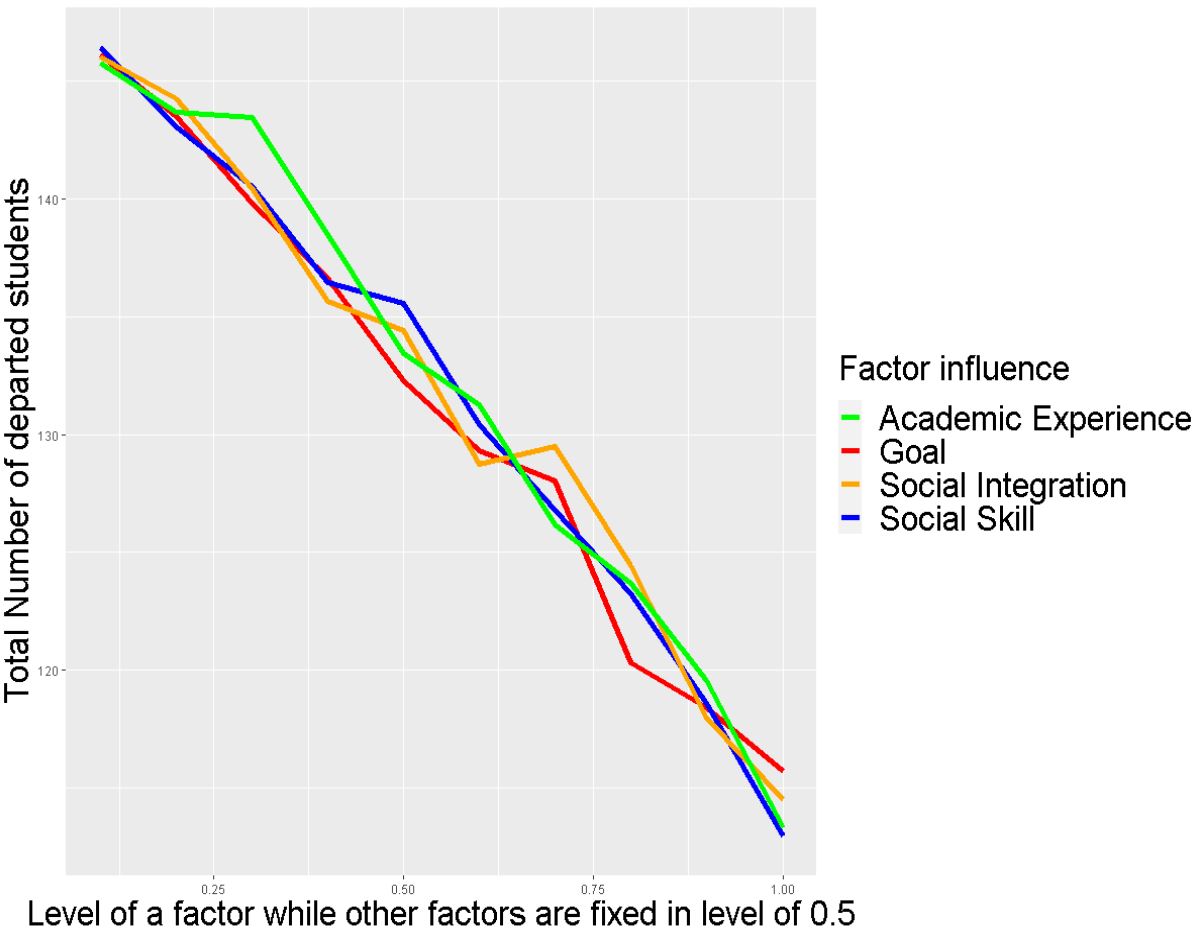}
    \caption{Number of departed students based on different levels of one factor while fixing the other factors.}
    \label{departed}
\end{subfigure}\hspace*{\fill}
\caption{Number of graduated and departed students based on different levels of one factor while fixing the other factors to the level of 0.5 during four years of post-secondary education.}
\label{graduated_departed}
\end{figure*}

\textbf{RUN Specifications.} 
To begin each run of the model, the user sets the initial number of agents, fraction of teachers, and the college attendance percentage as well as the initial values of the four non-cognitive factor variables. The model runs based on ticks (i.e,., each run ticks the model once). For simplification, in this work we only consider 4-year colleges or university programs (that we call "college" at the rest of this manuscript) for post-secondary education. Therefore, model runs for the total of four ticks to achieve the end of the fourth year at college. At the end of the fourth year, students graduate and leave the college.

The deaf agents can only decide whether to attend college or not in the first tick (that is, the first year). 
If they do, they move to one of empty spots at the college location and their color changes to yellow. Otherwise, they will remain at residential locations without any color change. We call the attended college agents "students". Students are deaf agents who not only attended college, but also are still persisting at college at the end of each tick (i.e., year). If attended agents decide to depart college during a tick, they move out of the college and move to a random spot in the residential locations. These agents who attended college but departed it are called "quitters". Quitters are shown as red human shapes located at residential locations. For simplicity, we assume that if students depart college, they do not attend it again.

In the second, third, and fourth ticks of the model, students can only decide whether to depart college or not. Just as in the first run, if they decide to depart, they will leave college and locate to a random spot of the residential locations. Their color will also turn to red. The screen of the model after running two ticks is shown in Figure \ref{during}.

During college years, students can randomly create links or connections (shown with gray lines in Figure \ref{during} and \ref{end}) with teachers or other students at the college. The total number of these links are used to adjust and update the level of social integration and academic experience factors of the student. The number of links that a student creates with teachers and other students will be used to evaluate the level of the academic experience and social integration factors of the student, respectively. The more links, the higher the level of the factor. The number of links that each student creates with other teachers and students can vary between zero to three and zero to eight, respectively. The minimum value of both social integration and academic experience of each student is considered to be $0.2$. Each created link with other students add $0.1$ to the minimum value of a student's social integration. Similarly, each created connection with a teacher at the college adds $0.1$ to the minimum value of the student's academic experience factor.

Similar to the first three ticks, during the fourth tick of the model, students can still persist in college. If they do, they have finished four years of college and therefore, they have completed the program have graduated. These students are called "graduates". Graduates, whose color change to blue, depart college at the end of the fourth year and move to one of the empty spots at the residential locations. Figure \ref{end} shows the model screen after the fourth tick.

\begin{figure*}
\begin{subfigure}{0.45\textwidth}
    \centering
    \includegraphics[width=\linewidth,height=7cm,width=7.5cm, scale=0.5]{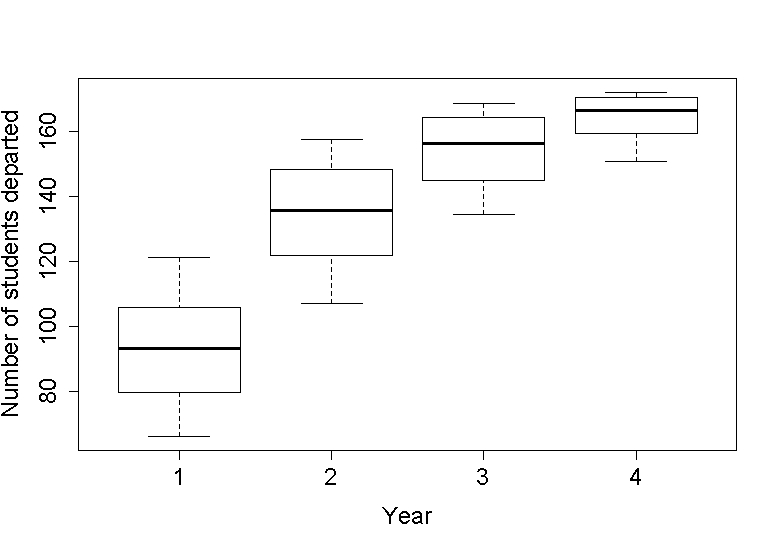}
    \caption{Boxplot of the results of the behavior space experiment for goal factor}
    \label{Boxplot_goal}
\end{subfigure}\hspace*{\fill}
\begin{subfigure}{0.45\textwidth}
    \centering
    \includegraphics[width=\linewidth,height=7cm,width=7.5cm, scale=0.5]{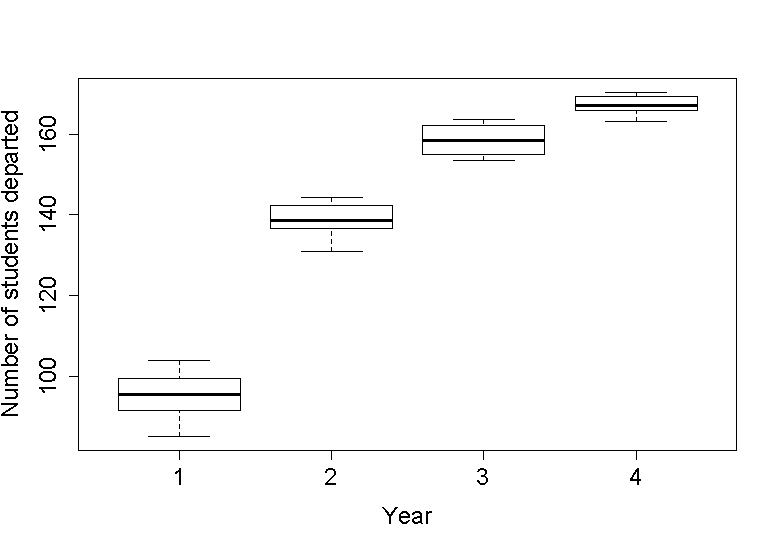}
    \caption{Boxplot of the results of the behavior space experiment for academic experience factor}
    \label{Boxplot_academic_experience}
\end{subfigure}\hspace*{\fill}

\medskip

\centering
\begin{subfigure}{0.45\textwidth}
    \centering
    \includegraphics[width=\linewidth,height=7cm,width=7.5cm, scale=0.5]{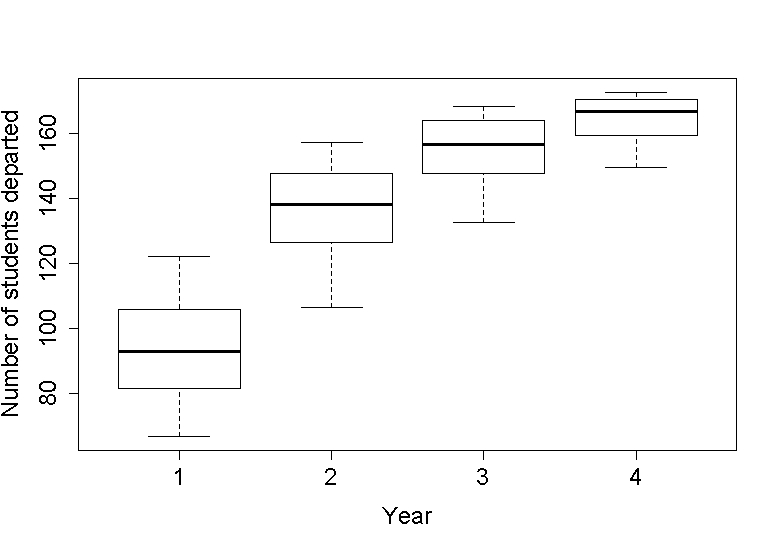}
    \caption{Boxplot of the results of the behavior space experiment for social skill factor}
    \label{Boxplot_social_skill}
\end{subfigure}\hspace*{\fill}
\begin{subfigure}{0.45\textwidth}
    \centering
    \includegraphics[width=\linewidth,height=7cm,width=7.5cm, scale=0.5]{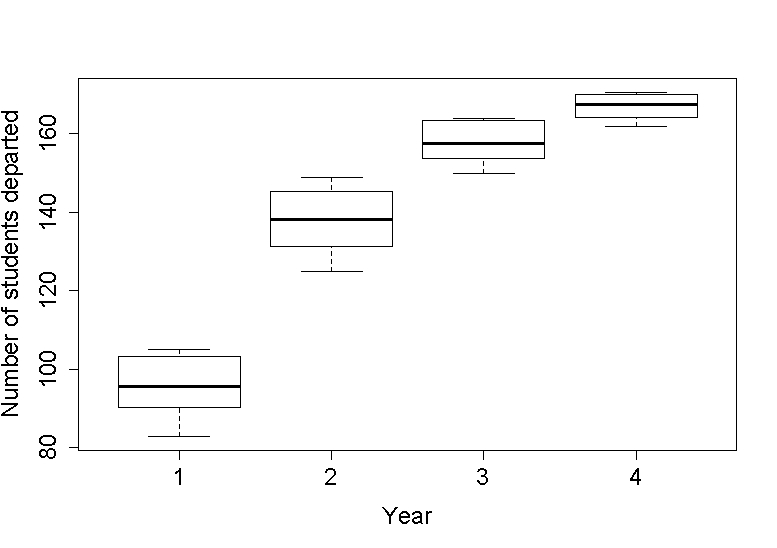}
    \caption{Boxplot of the results of the behavior space experiment for social integration factor}
    \label{Boxplot_social_integration}
\end{subfigure}\hspace*{\fill}
\caption{Boxplots of the results of the behavior space experiment for the four non-cognitive factors of goal, academic experience, social skill, and social integration. Each graph shows the result of the behavior search for one factor changing from interval of (0,1) with steps of $0.1$. To run the experiment for one factor, all other three factors are assigned the value of $0.5$.}
\label{boxplot}
\end{figure*}

\section{Results}\label{experiments and results}

\begin{figure*}[t]
\centering
\begin{subfigure}{\textwidth}
    \centering
    \includegraphics[width=0.65\linewidth,scale=0.5]{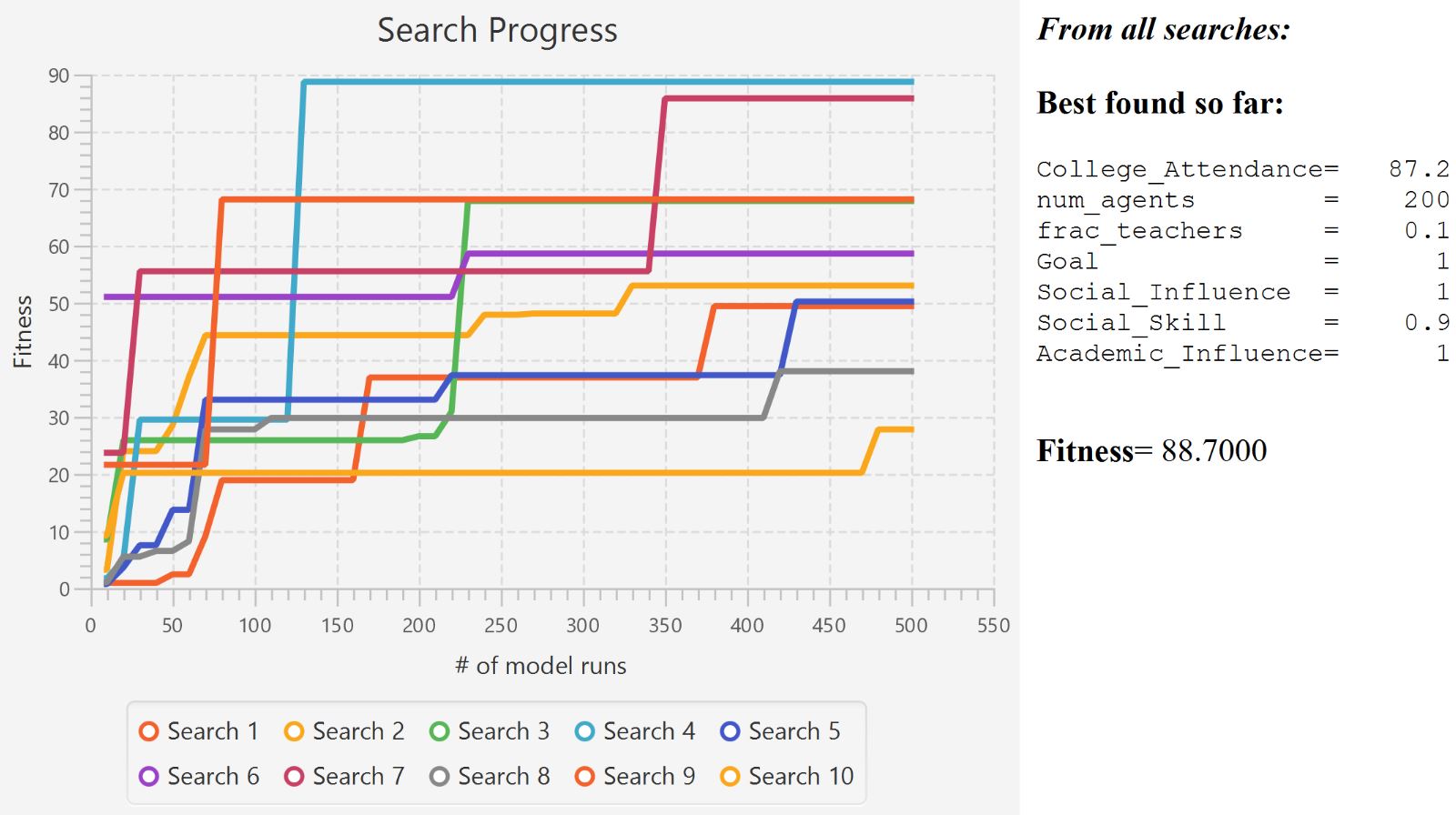}
    \caption{Plot of ten different searches for maximizing the total number of departed students. Maximized fitness is achieved with level 1 for goal, social skill, and social integration and level of 0.8 for academic integration.}
    \label{behavior_search_graduated}
\end{subfigure}\hspace*{\fill}
\vspace{0.5cm}
\medskip

\begin{subfigure}{\textwidth}
    \centering
    \includegraphics[width=0.65\linewidth,scale=0.5]{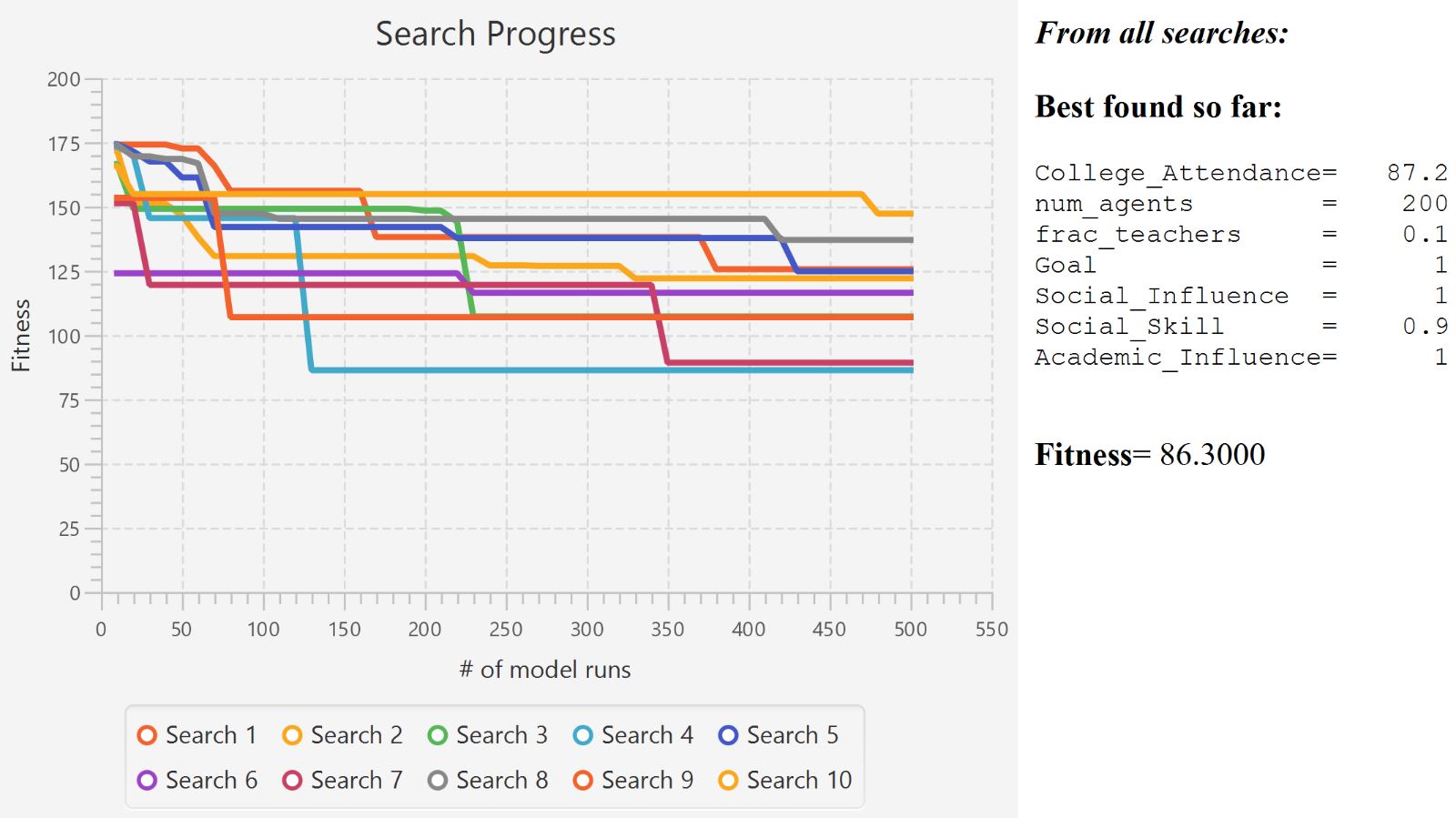}
    \caption{Plot of ten different searches for minimizing the total number of departed students. Minimized fitness is achieved with level 1 for goal, social skill, and social integration, and level of 0.8 for academic integration.}
    \label{behavior_search_departed}
\end{subfigure}\hspace*{\fill}
\caption{Result screens of running the Behavior Search tool on the proposed NetLogo model}
\label{BehaviorSearch_graduated_departed}
\end{figure*}

The implementation of this work is done using NetLogo 6.1.1 \cite{b31}. The results of replicating, repeating, and reproducing the results, with input assumptions for all runs are provided in Table \ref{runs}.

We performed four experiments using the behavior search tool of NetLogo to find the influence of each factor on the college persistence of deaf students for years one, two, three, and four. Persistence of four years at college is considered as graduating from college and receiving a 4-year college degree. In order to run each experiment, we considered one non-cognitive factor value changing in an interval of (0,1) with steps of 0.1 while the values of the other three factors are fixed at 0.5. The results of these four experiments are shown in Figure \ref{years}. As the figure shows, although increasing the level of a factor increases the persistence of students at post-secondary education, these four non-cognitive factors' impacts vary from one to another. In addition, as all four figures show, the black line showing the number of students who persisted one year at college is further away from the other lines for students who persisted two or more years at college. This shows that the majority of students depart post-secondary education during the first year of education. In other words, students who do not have clear goals and strategies, positive academic experience, strong social skills, and high levels of social integration are most likely to depart post-secondary education during the first year. If they persist after the first year, the chance of departing decreases with a high margin. 


In order to compare the number of graduated and departed students with different levels of non-cognitive factors together in two separate graphs, an experiment in which the level of all factors can change in an interval of (0.1,1) with steps of $0.1$ is performed. This enables us to extract a plot for the total number of graduated (i.e., all green lines of four\_year at college from Figure \ref{years} in one single plot) and departed students with different levels of factors. Figure \ref{graduated_departed} shows the graphs of the number of graduated and departed deaf students with fixed three non-cognitive factors at level 0.5 and varying one factor in an interval of (0.1,1) with steps of 0.1 during four years of post-secondary education. Different colored lines show the line plot for the assigned factor for them that is changing. As Figure \ref{departed} shows, although by increasing the level of factors the total number of departed students decreases, it still does not merge to zero. This indicates that despite the high levels of non-cognitive factors, some students may still depart college due to other factors, issues, or concerns. Furthermore, considering low levels of academic experience and social integration, the total number of departed students are noticeably more than other factors, while in contrast, with high levels of goal and social skill, the total number of departed students are noticeably decreased. This illustrates the importance of academic experience and social integration in low margins and strong goals and social skill in high margins when a student decides whether to depart college or not.

The results of the sensitivity analysis for each individual non-cognitive factor can be seen in Figure \ref{boxplot}. The +- 10\% range is illustrated using error bars while the boxplots represent the range of number of departed students in ten repetitions of the model. Although boxplot graphs of the four factors' influence are somewhat similar, it is apparent that the model is highly sensitive to the academic experience factor. In addition, as Figure \ref{years} illustrates, the number of departed students between year one to four is not linear, and the slope decreases as the year increases.

To maximize the number of graduated students as well as to minimize the total number of departed students, we performed a behavior search experiment using the proposed NetLogo model. Figure \ref{BehaviorSearch_graduated_departed} shows the results of the behavior search experiment for both aforementioned objectives. As the figure illustrates, the best value for both objectives is achieved by the value of 1 for goal, academic integration, and social integration factors, and value of 0.9 for social skill factor. The maximum number of graduated students as well as the minimum number of departed students based on the aforementioned factor values are $88.700$ and $86.3$, respectively. However, because we are counting the number of persons, both values will be rounded up to $89$ and $87$. These results show that almost half of deaf students decide to depart post-secondary education before graduation.

\section{conclusion}
In this work, we study the effects of four non-cognitive factors: having clear goals, social integration, social skills, and academic experience, on post-secondary persistence or retention of deaf students from an agent-based modeling (ABM) and simulation approach. To the best of our knowledge, we present the first ABM simulation for the aforementioned problem in order to simulate students retention behavior and discover the effects of non-cognitive factors in students persistence and departure decisions. Our results indicate that first year persistence at a 4-year post-secondary education (e.g., university, college) plays an integral role in student's persistence and graduation. In other words, if a student persists after the first year of a post-secondary education, the chances of student departure decreases with a high margin. In addition, the best persistent rate of the model is achieved by a social skill factor of 0.9 and other factors of 1. We believe that presenting and creating ABM brought significant benefits to studying deaf students' departure decisions during post-secondary education.






\begin{thebibliography}{00}

\bibitem{b1}   Boutin, D. L. (2008). Persistence in postsecondary environments of students with hearing impairments. Journal of Rehabilitation, 74(1), 25.

\bibitem{b2}   Aljohani, O. (2016). A Comprehensive Review of the Major Studies and Theoretical Models of Student Retention in Higher Education. Higher Education Studies, 6(2), 1-18.

\bibitem{b3}   National Council on Disability. (2003). People with disabilities and postsecondary education. Retrieved March 18, 2010, from http://www.ncd.gov/newsroom/publications/2003/education.htm

\bibitem{b4} Mamiseishvii, K., \& Koch, L. C. (2010). First-to-second year persistence of students with disabilities in postsecondary institutions in the united states. Rehabilitation Counseling Bulletin, 54(2), 93–105.

\bibitem{b5} Self-Assessments and other perceptions of successful adults who are deaf: An initial investigation. American Annals of the Deaf, 148(3), 243–250.

\bibitem{b6}   Garberoglio, C. L., Schoffstall, S., Cawthon, S., Bond, M., \& Ge, J. (2014). The role of self-beliefs in predicting postschool outcomes for deaf young adults. Journal of developmental and physical disabilities, 26(6), 667-688.

\bibitem{b7}   Test, D. W., Mazzotti, V. L., Mustian, A. L., Fowler, C. H., Kortering, L., \& Kohler, P. (2009). Evidence-based secondary transition predictors for improving postschool outcomes for students with disabilities. Career Development for Exceptional Individuals, 32(3), 160-181.

\bibitem{b8}   Garberoglio, C. L., Palmer, J. L., \& Cawthon, S. (2019). Undergraduate Enrollment of Deaf Students in the United States. Washington, DC: U.S. Department of Education, Office of Special Education Programs, National Deaf Center on Postsecondary Outcomes.

\bibitem{b9}   Garberoglio, C.L., Palmer, J.L., Cawthon, S., \& Sales, A. (2019a). Deaf People and Educational Attainment in the United States: 2019. Washington, DC: U.S. Department of Education, Office of Special Education Programs, National Deaf Center on Postsecondary Outcomes.

\bibitem{b10}	Stinson, M. S. \& Walter, G. G. (1997). Improving retention for deaf and hard of hearing students: What the research tells us. Journal of the American Deafness and Rehabilitation Association, 30, 14–23.

\bibitem{b11}	Wagner, M., Cadwallader, T., \& Marder, C. (2003). Life outside the classroom for youth with disabilities. Retrieved from https://nlts2.sri.com/reports/2003\_04-2

\bibitem{b12}	National Deaf Center on Postsecondary Outcomes (2019). Non-Cognitive Factors That Support Postsecondary Persistence in Deaf Students. Retrieved from https://www.nationaldeafcenter.org/file/non-cognitive-factors-support-postsecondary-persistence-deaf-students


\bibitem{b13}   Newman, L. A., Marschark, M., Shaver, D. M., \& Javitz, H. (2017). Course-Taking Effect on Postsecondary Enrollment of Deaf and Hard of Hearing Students. Exceptionality, 25(3), 170-185.

\bibitem{b14}   Harrison, R. H. (1987). Identifying factors that predict deaf students' academic success in college (Doctoral dissertation, Graduate School of Arts and Sciences, University of Pennsylvania).

\bibitem{b15}   Stinson, M. S., Scherer, M. J., \& Walter, G. G. (1987). Factors affecting persistence of deaf college students. Research in Higher Education, 27(3), 244-258.

\bibitem{b16} Cuculick, J., \& Kelly, R. (2003). Relating deaf students’ reading and language scores at college entry to their degree completion rates. American Annals of the Deaf, 148, 279–286.

\bibitem{b17}	Richardson, J. T. E., McLeod-Gallinger, J., McKee, B. G., \& Long, G. L. (1999). Approaches to studying in deaf and hearing students in higher education. Journal of Deaf Studies and Deaf Education, 5, 156–173.

\bibitem{b18}	Stinson, M. S., \& Walter, G. G. (1992). Persistence in college. In S. B. Foster \& G. G. Walter (Eds.), Deaf students in postsecondary education (pp. 43–64). New York, NY: Routledge.

\bibitem{b19} Assessing English literacy as a predictor of postschool outcomes in the lives of Deaf individuals. Journal of Deaf Studies and Deaf Education, 1–18. 9 Lang, H. (2002).

\bibitem{b20} Albertini, J. A., Kelly, R. R., \& Matchett, M. K. (2012). Personal factors that influence deaf college students’ academic success. Journal of Deaf Studies and Deaf Education, 17(1), 85–101.

\bibitem{b21} Danermark, B. (1995). Persistence and academic and social integration of hearing-impaired students in postsecondary education: A review of research. JADARA, 29(2), 20–33.

\bibitem{b22} Cawthon, S. W., Caemmerer, J. M., Dickson, D. M., Ocuto, O., Ge, J., \& Bond, M. (2015). Social skills as a predictor of postsecondary outcomes for individuals who are deaf. Applied Developmental Science, 19(1), 19–30.

\bibitem{b23}	Rogers, S., Muir, K., \& Evenson, C. R. (2003). Signs of resilience: Assets that support deaf adults’ success in bridging the deaf and hearing world. American Annals of the Deaf, 148(3), 222– 232.

\bibitem{b24}   Gilbert, N. (2019). Agent-based models (Vol. 153). Sage Publications, Incorporated.

\bibitem{b33}	Yousefi, N., Alaghband, M., \& Garibay, I. (2019). A Comprehensive Survey on Machine Learning Techniques and User Authentication Approaches for Credit Card Fraud Detection. arXiv preprint arXiv:1912.02629.

\bibitem{b34}	Alaghband, M., Yousefi, N., \& Garibay, I. (2020). FePh: An Annotated Facial Expression Dataset for the RWTH-PHOENIX-Weather 2014 Dataset. arXiv preprint arXiv:2003.08759.

\bibitem{b25}   Wilensky, U., \& Rand, W. (2015). An introduction to agent-based modeling: modeling natural, social, and engineered complex systems with NetLogo. Mit Press.

\bibitem{b26}   Epstein, J. M. (1999). Agent‐based computational models and generative social science. Complexity, 4(5), 41-60.

\bibitem{b27}   Bonabeau, E. (2002). Agent-based modeling: Methods and techniques for simulating human systems. Proceedings of the national academy of sciences, 99(suppl 3), 7280-7287.

\bibitem{b28}   Tinto, V. (1975) ‘Dropout From Higher Education: A Theoretical Synthesis of Recent Research.’ Review of Educational Research, 45, 89-125.

\bibitem{b29}   Tinto, V. (1993) Leaving College: Rethinking the Causes and Cures of Student Attrition, 2nd(ed.), Chicago: University of Chicago Press.

\bibitem{b30}   Kember, D. (1995) Open Learning Courses for Adults: A Model of Student Progress, Englewood Cliffs, New Jersey: Educational Technology Publications

\bibitem{b31}   Wilensky, U. (1999). NetLogo. http://ccl.northwestern.edu/netlogo/. Center for Connected Learning and Computer-Based Modeling, Northwestern University, Evanston, IL.

\bibitem{b32}	Stinson, M. S., Scherer, M. J., \& Walter, G G. (1987). Factors affecting persistence of deaf college students. Research in Higher Education, 27, 244–258.








\end{thebibliography}
\end{document}